\documentclass[12pt]{article}
\def\nabstar#1{\nabla\kern-0.5pt\smash{\raise 4.5pt\hbox{$\ast$}}
               \kern-4.5pt_{#1}}

\def\drvstar#1{\partial\kern-0.5pt\smash{\raise 4.5pt\hbox{$\ast$}}
               \kern-5.0pt_{#1}}

\def\newline{\relax\ifhmode\null\hfil\break\else\nonhmodeerr@\newline\fi}
\def\frac#1#2{{#1\over#2}}
\def\text#1{{\hbox{\rm #1}}}
\def\flushpar{{\par \noindent}}

\newcommand{\beq}{\begin{equation}}
\newcommand{\eeq}{\end{equation}}
\newcommand{\bea}{\begin{eqnarray}}
\newcommand{\eea}{\end{eqnarray}}

\def\BE{\begin{equation}}
\def\EE{\end{equation}}
\def\BA{\begin{eqnarray}}
\def\EA{\end{eqnarray}}
\def\BAN{\begin{eqnarray*}}
\def\EAN{\end{eqnarray*}}

\def\nn{\nonumber\\}

\def\gm5{\gamma_5}

\input epsf.sty
\newdimen\psfigsize
\def\psfigure#1 #2 #3 #4 #5{
    \begin{figure}[tbh]
      \begin{center}
      \vbox{
        \null\vskip-0.2in\hskip#2
        \epsfxsize=#1
        \epsfbox{#4}
        \vskip -0.3in
        \caption {#5 \label{#3}}
        \vskip 0.0 true in plus 0.3 true in
      }
      \end{center}
   \end{figure}
}
\begin{document}
\thispagestyle{empty}
\begin{flushright}
NTUTH-02-505H \\
UW-PT-02-26 \\
November 2002
\end{flushright}
\bigskip\bigskip\bigskip
\vskip 2.5truecm
\begin{center}
{\LARGE {Locality of optimal lattice domain-wall fermions}}
\end{center}
\vskip 1.0truecm
\centerline{Ting-Wai Chiu}
\vskip5mm
\centerline{Institute for Advanced Study, Princeton, NJ 08540, USA}
\smallskip
\centerline{and}
\smallskip
\centerline{Department of Physics, University of Washington}
\centerline{Seattle, WA 98195-1560, USA}
\smallskip
\centerline{and}
\smallskip
\centerline{Department of Physics, National Taiwan University}
\centerline{Taipei, Taiwan 106, Taiwan.}
\centerline{\it E-mail : twchiu@phys.ntu.edu.tw}
\vskip 1cm
\bigskip \nopagebreak \begin{abstract}
\noindent

It is shown that the effective 4D lattice Dirac operator of  
optimal lattice domain-wall fermions with finite $ N_s $ 
(in the fifth dimension) is exponentially local for 
sufficiently smooth gauge background. 



\end{abstract}
\vskip 1.5cm

\newpage\setcounter{page}1

Recently, it has been shown that one can construct a lattice 
domain-wall fermion action such that the effective 4D lattice 
Dirac operator preserves the chiral symmetry optimally for any 
given finite $ N_s $ in the fifth dimension \cite{Chiu:2002ir}.

Explicitly, the action of optimal lattice domain-wall fermions 
reads\footnote{In this paper, we suppress the lattice 
spacings ($ a $ and $ a_5 $), as well as the Dirac and color indices, 
which can be restored easily.}  
\bea
\label{eq:twc}
&& \hspace{-6mm}
{\cal A} = \sum_{s,s'=1}^{N_s} \sum_{x,x'}
\bar\psi(x,s)
 [ ( \omega_s D_w(x,x') + 1 ) \delta_{s,s'}
 + ( \omega_{s} D_w(x,x') - 1 ) P_{-} \delta_{s',s+1} \nn
&& \hspace{50mm}
 + ( \omega_{s} D_w(x,x') - 1  ) P_{+} \delta_{s',s-1} ] \psi(x',s') \nn
&& - \frac{m_q}{2m_0} \sum_{x,x'} \
     [ \bar\psi(x,1) ( \omega_{1} D_w(x,x') - 1 ) P_{+} \psi(x',N_s) \nn
&& \hspace{20mm}
  +\bar\psi(x,N_s) ( \omega_{N_s} D_w(x,x') - 1  ) P_{-} \psi(x',1) ] 
\eea
with boundary conditions 
\BAN
P_{+} \psi(x,0) = - \frac{m_q}{2m_0} P_{+} \psi(x,N_s), \hspace{6mm}
P_{-} \psi(x,N_s+1) = - \frac{m_q}{2m_0} P_{-} \psi(x,1) \ ,
\EAN
and the quark fields with bare mass $ m_q $ 
can be constructed from the left and right boundary modes
\BAN
q(x) &=& P_{-} \psi(x,1) + P_{+} \psi(x, N_s ) \\
\bar{q}(x) &=& \bar\psi(x,1) P_{+} + \bar\psi(x, N_s) P_{-} \ .
\EAN
Here $ D_w $ is the 4D Wilson-Dirac operator with a 
negative parameter ($ -m_0 $),
\BAN
D_w &=& \sum_{\mu=1}^4 \gamma_\mu t_\mu + W - m_0,
                       \hspace{6mm} m_0 \in (0,2) \\
t_\mu(x,x') &=& \frac{1}{2}[   U_\mu(x) \delta_{x',x+\mu}
                              - U_\mu^{\dagger}(x') \delta_{x',x-\mu} ] \\
W(x,x') &=& \sum_{\mu=1}^4 \frac{1}{2}[ 2 \delta_{x,x'}
                          - U_\mu(x) \delta_{x',x+\mu}
                          - U_\mu^{\dagger}(x') \delta_{x',x-\mu} ] \ , 
\EAN
and the weights  
\bea
\label{eq:omega}
\omega_s = \frac{1}{\lambda_{min}} \sqrt{ 1 - \kappa'^2 \mbox{sn}^2
           \left( v_s ; \kappa' \right) }, \hspace{6mm} s=1,\cdots,N_s 
\eea
where $ \mbox{sn}( v_s; \kappa' ) $ is the Jacobian elliptic function
with argument $ v_s $ (\ref{eq:vs}) and modulus
$ \kappa' = \sqrt{ 1 - \lambda_{min}^2 / \lambda_{max}^2 } $,
($ \lambda_{max}^2 $ and $ \lambda_{min}^2 $ are the maximum
and the minimum of the eigenvalues of $ H_w^2 $, $ H_w = \gamma_5 D_w $).
The weights $ \{ \omega_s \} $ are obtained from the roots
$ (u_s = \omega_s^{-2}, s=1,\cdots,N_s) $ of the equation
\BAN
\label{eq:delta_Z}
\delta_Z(u) =
     \left\{ \begin{array}{ll}
 1-\sqrt{u} R_Z^{(n,n)}(u)=0 \ ,   & \ N_s=2n+1 \nn
 1-\sqrt{u} R_Z^{(n-1,n)}(u)=0 \ , & \ N_s=2n   \nn
             \end{array} \right.
\EAN
The argument $ v_s $ in (\ref{eq:omega}) is    
\bea
\label{eq:vs}
v_s &=& (-1)^{s-1} M \ \mbox{sn}^{-1}
    \left( \sqrt{\frac{1+3\lambda}{(1+\lambda)^3}}; \sqrt{1-\lambda^2} \right)
  + \left[ \frac{s}{2} \right] \frac{2K'}{N_s}
\eea
where
\bea
\label{eq:lambda}
\lambda &=&
\prod_{l=1}^{N_s}
\frac{\Theta^2 \left(\frac{2lK'}{N_s};\kappa' \right)}
     {\Theta^2 \left(\frac{(2l-1)K'}{N_s};\kappa' \right)} \ , \\
\label{eq:M}
M &=&
\prod_{l=1}^{[\frac{N_s}{2}]}
\frac{\mbox{sn}^2 \left(\frac{(2l-1)K'}{N_s};\kappa' \right) }
{ \mbox{sn}^2 \left(\frac{2lK'}{N_s};\kappa' \right) } \ ,
\eea
$ K' $ is the complete elliptic integral of the first kind with
modulus $ \kappa' $, and $ \Theta $ is the elliptic theta function.
Note that $ \lambda_{max}^{-1} \le \omega_s \le \lambda_{min}^{-1} $, 
since $ \mbox{sn}^2(;) \le 1 $.

After adding pseudofermions (Pauli-Villars fields)
with fixed mass $ m_q = 2 m_0 $ to the action (\ref{eq:twc}), 
one can derive the effective 4D lattice Dirac operator for the 
internal quark loops 
\bea
\label{eq:D}
D(m_q) = r( D_c + m_q )( 1 + r D_c )^{-1}, \hspace{6mm} r = \frac{1}{2m_0}
\eea
where 
\bea
\label{eq:Dc}
r D_c &=& \frac{1+\gamma_5 {\cal S}_{opt}}{1-\gamma_5 {\cal S}_{opt}} \\
{\cal S}_{opt}  
&=& \left\{ \begin{array}{ll}
           H_w R_Z^{(n,n)}(H_w^2),   &  N_s = 2n+1   \nn
           H_w R_Z^{(n-1,n)}(H_w^2), &  N_s = 2n,    \nn
          \end{array} \right.  
\label{eq:S_opt}
\eea
and $ R_Z(H_w^2) $ is the Zolotarev optimal rational approximation 
for the inverse square root of $ H_w^2 $,
\bea
\label{eq:rz_nn}
R^{(n,n)}_Z(H_w^2) = \frac{d_0}{\lambda_{min}}
\prod_{l=1}^{n} \frac{ 1+ h_w^2/c_{2l} }{ 1+ h_w^2/c_{2l-1} }
\eea
and
\bea
\label{eq:rz_n1n}
R^{(n-1,n)}_Z(H_w^2) = \frac{d'_0}{\lambda_{min}}
\frac{ \prod_{l=1}^{n-1} ( 1+ h_w^2/c'_{2l} ) }
     { \prod_{l=1}^{n} ( 1+ h_w^2/c'_{2l-1} ) }
\eea
in which $ h_w^2 = H_w^2/\lambda_{min}^2 $,   
and the coefficients $ d_0 $, $ d'_0 $, $ c_l $ and $ c'_l $
are expressed in terms of elliptic functions \cite{Akhiezer:1990} 
depending on $ N_s $ and $ \lambda_{max}^2 / \lambda_{min}^2 $.

In the limit $ N_s \to \infty $, ${\cal S}_{opt} \to H_w (H_w^2)^{-1/2}$,
thus $ r D_c $ (\ref{eq:Dc}) becomes chirally symmetric. 
Therefore, in the massless limit ($m_q=0$) and $ N_s \to \infty $,
$ D $ (\ref{eq:D}) is exactly equal to the overlap Dirac operator
\cite{Neuberger:1998fp,Narayanan:1995gw}, and
satisfies the Ginsparg-Wilson relation \cite{Ginsparg:1981bj}
\BAN
D \gamma_5 + \gamma_5 D = 2 D \gamma_5 D \ .
\EAN
This implies that the effective 4D Dirac operator of optimal DWF
is exponentially local for sufficiently smooth gauge background,
and is topologically-proper (i.e., with the correct index and
axial anomaly), similar to the case of overlap Dirac operator.
At finite $ N_s $, the effective 4D lattice Dirac operator
is exactly equal to the overlap Dirac operator with $ (H_w)^{-1/2} $
approximated by Zolotarev optimal rational polynomial.

The valence quark propagator coupling to physical hadrons is  
\BAN
\label{eq:valence_q}
( D_c + m_q )^{-1} = r ( 1 - r m_q )^{-1} [ D^{-1}(m_q) - 1 ] \ ,
\EAN
where $ D^{-1}(m_q) $ can be computed via the 5-dimensional
lattice Dirac operator of optimal DWF. 

In this paper, we show that the effective 4D lattice Dirac 
operator (\ref{eq:D}) is exponentially local for sufficiently 
smooth gauge configurations. Since Eq. (\ref{eq:D}) can be rewritten as 
\bea
\label{eq:Dmq}
D(m_q) = r m_q + ( 1 - r m_q ) D \ , 
\eea
where
\bea
\label{eq:D_0}
D = r D_c ( 1 + r D_c )^{-1} = \frac{1}{2}( 1 + \gamma_5 {\cal S}_{opt} ) \ ,
\eea
thus it suffices to assert the exponential locality of $ D $ (\ref{eq:D_0}).  

First, we rewrite (\ref{eq:rz_nn}) and (\ref{eq:rz_n1n}) in partial 
fractions
\bea
\label{eq:RZnn}
R^{(n,n)}_Z(H_w^2) &=& ( 1 + h_w^2/c_{2n} )
                    \sum_{l=1}^{n} \frac{ b_l }{ 1+ h_w^2/c_{2l-1} } \\
\label{eq:RZn1n}
R^{(n-1,n)}_Z(H_w^2) &=& 
                     \sum_{l=1}^{n} \frac{ b_l }{ 1+ h_w^2/c_{2l-1} } 
\eea
where
\bea
\label{eq:cl}
c_l &=& \frac{\mbox{sn}^2(\frac{lK'}{N_s}; \kappa' ) }
             {1-\mbox{sn}^2(\frac{lK'}{N_s}; \kappa' )} \ , \\                  
\label{eq:bl}
b_l &=& \frac{1}{\lambda_{min}} \frac{2 \lambda }{1+ \lambda} \frac{1}{M}
        \frac{ \prod_{i=1}^{n-1} ( 1 - c_{2l-1}/c_{2i} ) }
             { \prod_{i=1, i \ne l}^{n} ( 1 - c_{2l-1}/c_{2i-1} ) } \ ,
\eea
$ \lambda $ and $ M $ are defined in (\ref{eq:lambda}) and 
(\ref{eq:M}) respectively.

Therefore, it suffices to examine the locality of each term 
in (\ref{eq:RZnn}) and (\ref{eq:RZn1n}), by expanding each 
fraction in terms of orthogonal polynomials, similar to 
the approach used in \cite{Hernandez:1998et}.
Here I expand each fraction in term
of Chebycheff polynomials of the first kind
\bea
\label{eq:cheby}
\frac{1}{1 + h_w^2/c_{2l-1}}= 
\frac{2 c_{2l-1}}{(b-1) \sinh \theta_l } 
\left( 1 + 2 \sum_{k=1}^{\infty} e^{-k \theta_l } \ T_k (z) \right)
\eea
where 
\bea
b &=& \frac{\lambda_{max}^2 }{\lambda_{min}^2} \ , \\
\label{eq:theta}
\theta_l &=& \cosh^{-1} \left(\frac{b+1+2 c_{2l-1}}{b-1} \right) \ , \\
z &=& \frac{ b + 1 - 2h_w^2 }{ b - 1} \ .  
\eea 
The expansion (\ref{eq:cheby}) is valid provided that $ \theta_l \ne 0 $ 
(i.e., $ \lambda_{min}^2 > 0 $), which is ensured by any gauge 
configuration satisfying \cite{Neuberger:1999pz}
\bea
\label{eq:U_bound}
\forall_{plaquette} || 1-U(p) || < \epsilon < \frac{1-|1-m_0|^2}{6(2+\sqrt{2})}
\eea
since 
\bea
\label{eq:Hw_bound}
\left( \sqrt{1-6(2+\sqrt{2})\epsilon} - | 1 - m_0 | \right)^2 
\le \lambda_{min}^2 < \lambda_{max}^2 \le ( | 4 - m_0 | + 4 )^2 \ .
\eea
 
Because $ h_w $ only has nearest neighbor coupling, it follows that 
$ h_w^2(x,y) = 0 $ for all separations bigger than two lattice spacings, 
i.e., $ \sum_{\mu} | x_\mu - y_\mu | > 2 a $. Thus, for the Chebycheff 
polynomial $ T_k(z) $ of degree $ k $, 
$ [T_k(z)](x,y) = 0 $ for $ \sum_{\mu} | x_\mu - y_\mu | > 2 k a $.
Then, (\ref{eq:cheby}) gives  
\bea
\label{eq:cheby_1}
\left(\frac{1}{1 + h_w^2/c_{2l-1}}\right)(x,y)= 
\frac{2 c_{2l-1}}{(b-1) \sinh \theta_l } 
\sum_{k=m}^{\infty} e^{-k \theta_l } \ [T_k(z)](x,y) \ , 
\eea
where $ m = \sum_{\mu} | x_\mu - y_\mu | / 2 a \ge 1 $. 
Using the triangle inequality and the property $ || T_k(z) || \le 1 $, 
we turn Eq. (\ref{eq:cheby_1}) into the inequality  
\BAN
\left|\left| \left(\frac{1}{1 + h_w^2/c_{2l-1}}\right) (x,y) \right|\right| 
&\le& 
\frac{2 c_{2l-1}}{(b-1) \sinh \theta_l } 
\frac{e^{-m \theta_l}}{(1-e^{-\theta_l})} \ , 
\EAN
and it follows that    
\bea
\label{eq:bound_th}
&& \left|\left|\left(
\sum_{l=1}^n \frac{b_l}{1+h_w^2/c_{2l-1}}\right)(x,y) \right|\right| \nn 
&\le& 
\frac{2}{b-1} \sum_{l=1}^{n}  
\frac{b_l c_{2l-1}}{ \sinh \theta_l } 
\frac{1}{(1-e^{-\theta_l})}
\exp \left(-\frac{\theta_l}{2a} \sum_{\mu} | x_\mu - y_\mu | \right) \nn  
&<& 
\frac{2}{b-1} \left( \sum_{l=1}^{n}  
\frac{b_l c_{2l-1}}{ \sinh \theta_l } 
\frac{1}{(1-e^{-\theta_l})} \right) 
\exp \left(-\frac{\theta_1}{2a} \sum_{\mu} | x_\mu - y_\mu | \right) \ ,   
\eea
where the last inequality is due to the property   
($ \theta_1 < \theta_2 < \cdots < \theta_n $) from (\ref{eq:theta}), 
since ($ 0 < c_1 < c_3 < \cdots < c_{2n-1} $) from (\ref{eq:cl}).
From (\ref{eq:D_0})-(\ref{eq:RZn1n}), and 
(\ref{eq:bound_th}), it follows that $ D $ (\ref{eq:D_0})
is exponentially-local, and so is $ D(m_q) $ (\ref{eq:Dmq}).

Obviously, the above analysis also ensures that the overlap Dirac 
operator with $ (H_w^2)^{-1/2} $ approximated by Zolotarev optimal 
rational polynomial is exponentially-local. Note that, in practice, 
the least upper bound of (\ref{eq:bound_th}) is much smaller 
than that on the r.h.s. of (\ref{eq:bound_th}).     
This implies that $ D(x,y) $ is still exponentially local even for
rough gauge configurations bounded by an $ \epsilon $ much larger  
than its least upper bound in (\ref{eq:U_bound}). 

Further, it is straightforward to assert the differentibility 
of $ D $ (\ref{eq:D_0}) with respect to the gauge field, 
as well as the exponential locality of $ D $ in the presence 
of an isolated near-zero modes of $ H_w^2 $, thus the details can be 
omitted. 
   
Finally, it is instructive to observe that the decay constant 
$ \theta_1 $ in (\ref{eq:bound_th}) is a monotonic decreasing    
function of $ N_s $. In the limit $ N_s \to \infty $, 
$ c_1 \to 0 $, and $ \theta_1 $ becomes the minimum,     
\BAN
\theta = \cosh^{-1} \left(\frac{b+1}{b-1} \right)  
= \cosh^{-1} \left( 
\frac{\lambda_{max}^2+\lambda_{min}^2}
{\lambda_{max}^2-\lambda_{min}^2} \right) \ . 
\EAN
In other words, at any finite $ N_s $,  
$ D(x,y) $ is more localized than itself at $ N_s = \infty $. 
In the limit $ \omega_s = 1 $, the effective 4D lattice Dirac 
operator is equivalent to the overlap Dirac operator with polar
approximation, and $ c_1 $ is replaced with $ d_1=\tan^2(\pi/2 N_s) $, 
thus its decay constant ($ \theta_1 $) is always smaller 
(i.e., less localized) than that of the optimal DWF, 
for any finite $ N_s $. 

For the conventional DWF with open boundary conditions, 
it has been estimated \cite{Kikukawa:1999dk} that the effective 
4D lattice Dirac operator at $ N_s = \infty $ behaves as 
\BAN 
|| D_{\mbox{eff}}(x,y) || < \ \sim 
\left( \sum_{\mu} | x_\mu - y_\mu | \right)    
\exp \left(-\frac{\tilde\theta}{2a} \sum_{\mu} |x_\mu - y_\mu| \right) \ .   
\EAN
However, for finite $ N_s $, the decay constant of 
$ D_{\mbox{eff}}(x,y) $ has not been obtained. 
Note that the factor ($ \sum_{\mu} |x_\mu - y_\mu| $) in   
$ D_{\mbox{eff}}(x,y) $ makes it decay slower than purely exponential decay. 
Nevertheless, this is only an artifact due to the 
expansion \cite{Kikukawa:1999dk} 
in terms of Chebycheff polynomials of the second kind.
If one expands the fraction in terms of 
Chebycheff polynomials of the first kind as (\ref{eq:cheby}), 
then one obtains a purely exponential bound,    
\BAN 
|| D_{\mbox{eff}}(x,y) || < \ \sim 
\exp \left(-\frac{\tilde\theta}{2a} \sum_{\mu} | x_\mu - y_\mu | \right) \ .   
\EAN

\vfill\eject
  
\bigskip
\bigskip
\flushpar
{\bf Acknowledgement}
\bigskip


This work was supported in part by National Science Council,
ROC, under the grant numbers NSC91-2112-M002-025 and NSC-40004F, and 
it was completed during my stay at Princeton Institute for Advanced Study. 
I am grateful to The Institute for Advanced Study, in particular Steve Adler, 
for support and kind hospitality. 
I also thank Herbert Neuberger for discussions.

\end{document}